\def\cm{{\rm\thinspace cm}}

\def\erg{{\rm\thinspace erg}}

\def\K{{\rm\thinspace K}}
\def\keV{{\rm\thinspace keV}}
\def\km{{\rm\thinspace km}}
\def\kpc{{\rm\thinspace kpc}}

\def\Zsun{\hbox{$\rm\thinspace Z_{\odot}$}}

\def\Msun{\hbox{$\rm\thinspace M_{\odot}$}}

\def\pc{{\rm\thinspace pc}}

\def\s{{\rm\thinspace s}}
\def\yr{{\rm\thinspace yr}}

\def\ergpcmsqps{\hbox{$\erg\cm^{-2}\s^{-1}\,$}}

\def\ergps{\hbox{$\erg\s^{-1}\,$}}

\def\kmps{\hbox{$\km\s^{-1}\,$}}

\def\Mpcc{{\rm\thinspace Mpc^{-3}}}

\def\Msunpyr{\hbox{$\Msun\yr^{-1}\,$}}
\def\pcm{\hbox{$\cm^{-3}\,$}}

\def\pcmsq{\hbox{$\cm^{-2}\,$}}

\documentclass[usegraphicx]{mn2e}

\usepackage{fixltx2e}
\usepackage{times}

\include{defn}
\begin{document}

\title[X-rays from NGC\,1277]{X-ray emission from the Ultramassive
  Black Hole candidate NGC\,1277: implications and speculation on its
  origin} \author[Fabian et al]
{\parbox[]{6.5in}{{A.C. Fabian$^1\thanks{E-mail: acf@ast.cam.ac.uk}$,
      J.S. Sanders$^2$, M. Haehnelt$^1$, M.J.Rees$^1$ and
      J.M. Miller$^3$
    }\\
    \footnotesize
    $^1$ Institute of Astronomy, Madingley Road, Cambridge CB3 0HA\\
    $^2$ Max-Planck-Institute f\"ur extraterrestrische Physik, 85748
    Garching, Germany\\
    $^3$ Department of Astronomy, University of Michigan, 500 Church
    St., Ann Arbor, MI48109, USA\\
  } }

\maketitle
  
\begin{abstract}
  We study the X-ray emission from NGC\,1277, a galaxy in the core of
  the Perseus cluster, for which van den Bosch et al. have recently
  claimed the presence of an UltraMassive Black Hole (UMBH) of mass
  $1.7\times 10^{10}\Msun$, unless the IMF of the stars in the stellar
  bulge is extremely bottom heavy. The X-rays originate in a power-law
  component of luminosity $1.3\times 10^{40}\ergps$ embedded in a 1
  keV thermal minicorona which has a half-light radius of about
  0.36~kpc, typical of many early-type galaxies in rich clusters of
  galaxies. If Bondi accretion operated onto the UMBH from the
  minicorona with a radiative efficiency of 10 per cent, then the
  object would appear as a quasar with luminosity $10^{46}\ergps$, a
  factor of almost $10^6$ times higher than observed. The accretion
  flow must be highly radiatively inefficient, similar to past results
  on M87 and NGC3115. The UMBH in NGC\,1277 is definitely not
  undergoing any significant growth at the present epoch. We note that
  there are 3 UMBH candidates in the Perseus cluster and that the 
  inferred present mean mass density in UMBH could be
  $10^5\Msun\Mpcc$, which is 20 to 30 per cent of the estimated mean
  mass density of all black holes. We speculate on the implied growth
  of UMBH and their hosts, and discuss the possibiity that extreme AGN
  feedback could make all UMBH host galaxies have low stellar masses
  at redshifts around 3. Only those which end up at the centres of
  groups and clusters later accrete large stellar envelopes and become
  Brightest Cluster Galaxies.  NGC\,1277 and the other Perseus core
  UMBH, NGC\,1270, have not however been able to gather more stars or
  gas owing to their rapid orbital motion in the cluster core. 
\end{abstract}


\section{Introduction}

The discovery of UltraMassive Black Holes (UMBH), with masses above
$10^{10}\Msun$ by McConnell et al (2011, 2012) has challenged our
understanding of black hole growth. Their objects lie at the centres
of Brightest Cluster Galaxies, which have the highest stellar masses
known. The mystery has now deepened by the recent claim of a UMBH in
the lenticular galaxy NGC\,1277 (van den Bosch et al 2012). The
estimated black
hole mass of $1.7\times 10^{10}\Msun$ corresponds to 14 per cent of
the total stellar mass of that galaxy, much larger than the 0.1 per
cent found in a normal massive galaxy. Moreover, van den Bosch et al
(2012) imply that there may be many more such objects, based on a
Table including 5 other galaxies with similar properties. All of the
known examples lie within about 100~Mpc from us.

Presumably they were among the most luminous quasars when they were
growing but are certainly not quasars now. NGC\,1277 lies in the core
of the Perseus cluster where we have amassed a very deep X-ray
exposure with Chandra (Fabian et al 2006; 2011). It appears as a weak
compact source with properties similar to other early-type galaxies in
the cores of clusters, showing a power-law continuum in an extended
thermal minicorona (Sun et al 2007; Santra et al 2007). We present
here a re-analysis of the X-ray data. The minicorona of 10 million K
gas lies within the Bondi radius so should be accreting. If the
accretion flow were radiatively efficient with a radiative efficiency of 10 per
cent then the luminosity should exceed $10^{46}\ergps$, i.e. it should
be a quasar. It clearly is not, since the power-law component has an
X-ray luminosity of only $1.3\times 10^{40}\ergps$. We discuss possible
ways in which accretion may be reduced. We then briefly look at
NGC\,1270 which is another object from the Table of UMBH candidates of
van den Bosch
(2012) and lies also in the Perseus core and has roughly similar
properties.

\begin{figure*}
  \centering
  \includegraphics[width=\textwidth,angle=0]{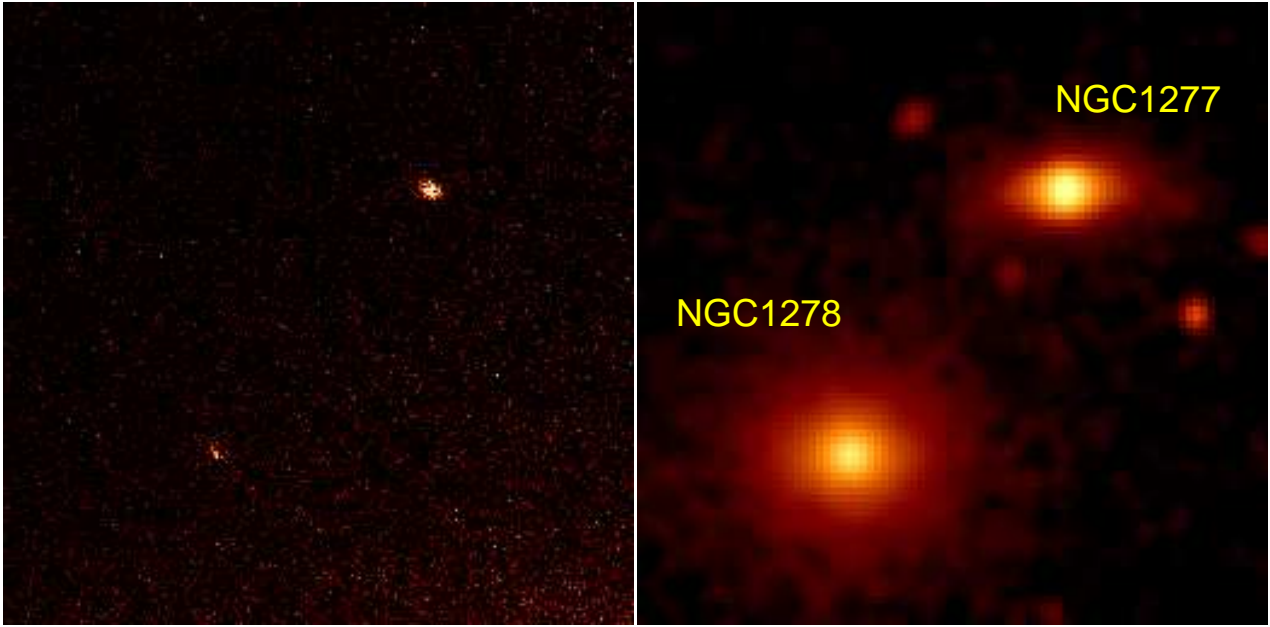}
  \caption{Left: Chandra image of NGC\,1277 and NGC\,1278 in the
    0.3-7~keV band. The two
    sources are separated by 47 arcsec. Right: IPAC-2MASS image of the same
  area of Sky. }
\end{figure*}

We  note the surprisingly large mean black hole mass density implied
by the newly discovered UMBH candidates.
We then consider the possibility in which extreme Active Galactic Nucleus  
feedback (Silk \& Rees 1998; Fabian 2012) during the growth phase of a UMBH
pushes all surrounding gas away, except in the directions along 
the filaments fuelling the black hole accretion. Star formation would
then be inhibited close to the UMBH and when growth ceases around
redshift 3, all UMBH are surrounded by whatever compact stellar bulge has
survived from the earliest growth phases. Those UMBH which now lie at the
centre of a cluster or group of galaxies will have since been able to
accrete a large stellar envelope due to cannibalism, cooling flows
etc and now appear as a BCG. Those orbiting in the core of a cluster, 
like NGC\,1277, will have been unable to accrete much of a  stellar
envelope and so have a large fraction of their mass in the central
black hole. 

We concentrate on the UMBH interpretation of NGC\,1277, but briefly
note and discuss the possible influence of a bottom-heavy stellar mass
function, which may reduce the black hole mass required by the
observations.

\section{The Chandra data}

The core of the Perseus cluster has been observed several times with Chandra
ACIS-S, centred on the Brightest Cluster Galaxy NGC\,1275 Fabian et al
(2006, 2012).  4345
background-subtracted counts are detected in a 3 arcsec radius
aperture on NGC\,1277 (Fig.~1), 3.75 arcmin to the North of NGC\,1275,
with a total effective exposure time of 698,800~s. The source flux in the
0.5--7~keV band is $4.1\times 10^{-14}\ergpcmsqps$.

\begin{figure}
  \centering
  \includegraphics[width=\columnwidth,angle=0]{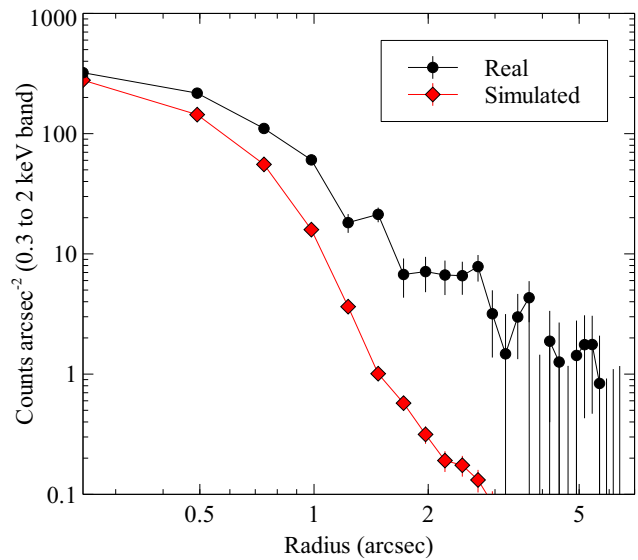}
  \caption{Radial profile of the image of NGC\,1277 (black points)
    compared with a simulation of the PSF carried out with the Chandra tool {\sc
      MARX} (red points).}
\end{figure}

Since the source is about 3 arcmin from the aimpoint of the telescope,
the point spread function is slightly worse than on axis. Fitting the
profile of the source by a gaussian gives a width, $\sigma$ of about 1
arcsec, which is slightly larger than the nominal width for this
off-axis distance. Using a simulation with the Chandra tool {\sc
  MARX}, we can compare it to the observed profile of the source
(Fig.~2). The Chandra data can be fitted by the simulated point source
of gaussian width (standard deviation) $\sigma_1=0.41$~arcsec
plus an extended part of intrinsic gaussian width
$\sigma_2=0.8$~arcsec. The ratio of the flux in the two components is
$\sim 1.6$. Adopting a cluster redshift of 0.018 the angular scale is
360~pc per arcsec, so the extended component has a half-light radius
of approximately 0.36~kpc.

The spectrum (Fig.~3) is characteristic of minicoronae (Vikhlinin et
al 2001; Sun et al 2007). It is well-modelled by thermal gas plus a
power-law continuum ($\chi^2/{\rm dof}=103.7/127$, the thermal
  model alone gives $144.1/129$ and the power-law alone gives
  $150.5/129$). The peak just below 1~keV is due to Fe-L emission
lines from the hot gas. We obtain a temperature for the gas of
$kT=1.02\pm{0.06}\keV$. The best-fitting metal abundance is 0.6 but is
poorly constrained provided that it is at least 0.05 (the thermal
continuum is degenerate with the power-law). Galactic absorption of
$1.28\times 10^{21}\pcmsq$ is included, which is typical for the core
of the Perseus cluster (Fabian et al 2006). The power-law component
has a photon index $\Gamma=1.93^{+0.20}_{-0.19},$ if the metallicity
is fixed at $1\Zsun$ (Anders \& Grevesse 1982), and the
absorption-corrected 0.5--7~keV luminosity is $1.3\times
10^{40}\ergps$. The normalization of the power-law component
  correlates with both $\Gamma$ and metal abundance $Z$. Provided that
  $Z>0.5\Zsun$ (i.e. at least comparable to the ICM) then the
  uncertainty of the normalization is less than 30 per cent.  These
results are similar to those obtained previously (Sun et al 2007). The
ratio of the flux of the power-law to the thermal power-law component
is about 2, in fair agreement with the power-law arising from an
unresolved central point source and the thermal emission being
extended, as expected. A good fit is also obtained with two thermal
components, at 0.96 and 4.6\,keV.

\begin{figure}
  \centering
  \includegraphics[width=0.62\columnwidth,angle=-90]{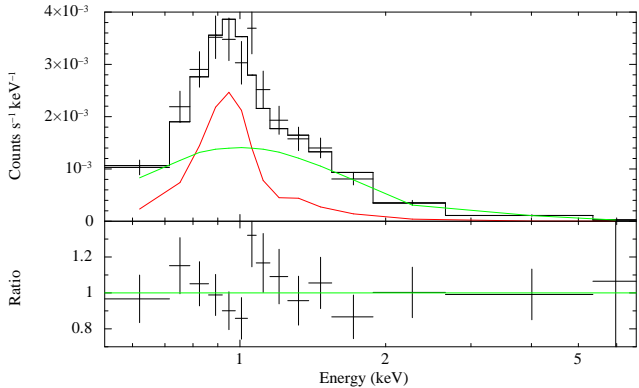}
  \caption{Chandra X-ray spectrum of NGC\,1277, fitted with Galactic
    absorption applied to an APEC
    thermal model (red) and a power-law continuum (green).}
\end{figure}
\begin{figure}
  \centering
  \includegraphics[width=\columnwidth,angle=0]{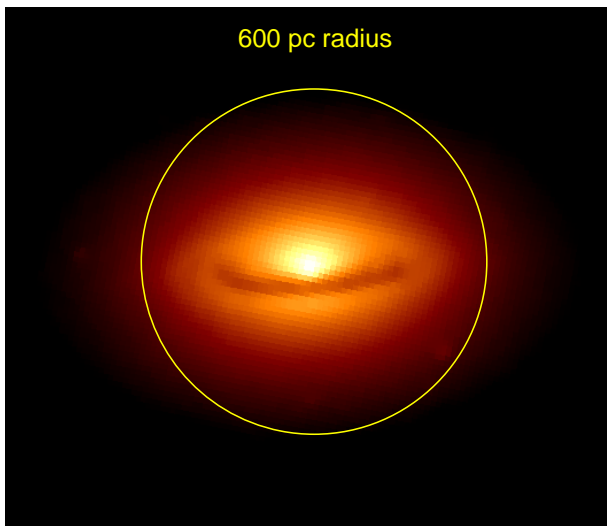}
  \caption{Preview image of HST-ACS image of NGC\,1277 taken through the
    F550M filter. A deep dust lane is clearly visible. }
\end{figure}

\section{The accretion power}
Modelling the thermal component as a sphere of uniform density and
radius $r\kpc$ gives a density of $0.023r^{-1.5}\pcm$.  The half-light
radius of sphere is 0.36~kpc if $r=0.6$. The Bondi radius $r_{\rm
  B}=GM/c^{2}$ for a black hole of mass $1.7\times 10^{10}\Msun$ is
$840\pc$ for gas at $10^7\K$, which is slightly larger than the radius
at which most of the thermal gas is found. Assuming that the gas is
flowing inward at its sound speed, at a radius of about 0.6~kpc, we
obtain an approximate accretion rate,
$$\dot M \sim 4 \pi \lambda r^2 n m_{\rm p} c_{\rm s}= 1.4 \Msunpyr,$$
which gives an accretion luminosity of
$$L_{\rm B}\sim 10^{46} \eta_{0.1} \ergps,$$ where we use 0.25 for the factor
$\lambda$, which is appropriate for Bondi accretion of an adiabatic
gas.  The radiative efficiency of accretion is $0.1\eta_{0.1}$, the
gas density $n$, the sound speed $c_{\rm s}$. $M$ and $m_{\rm p}$ are
the mass of the black hole and the proton, respectively.

There is then a mismatch between the predicted accretion luminosity and the
observed luminosity of the power-law source, by a factor of nearly $10^6$. Some
of this could be in bolometric corrections, but a factor of more than
$10^5$ will remain. We see no obvious disturbance in surrounding
intracluster gas so there is no major mechanical energy loss through
winds or jets. The overall accretion efficiency must be $10^{-6}$ or less. 
Inefficient accretion flows have been inferred before for Sgr A* at
the centre of our Galaxy (Rees 1983; Narayan \& Yi 1994). The flow can
become convectively unstable at low accretion rates leading to a
convectively dominated accretion flow (Narayan et al 2000; Quatert \&
Gruzinov 2000). Mismatches with supermassive black holes have  been
observed before, from M87 (Di Matteo et al 2003) and NGC\,3115 (Wong et al
2011) where $\eta<10^{-7}$. The most massive black
holes provide therefore stringent tests of accretion theory.

We note that there is a very deep dust lane (Fig.~4) in the centre of NGC\,1277,
with a radius of $\sim 300$~pc. This presumably indicates that matter
is orbiting the black hole there, with a velocity of at least $480\kmps$, if
the black hole mass is $1.7\times 10^{10}\Msun.$ Any cold accretion flow
may be stalled at the location of the dust lane: thermal gas at a 
temperature of 1~keV will
however be in a thick atmosphere. NGC\,1277 is listed as a radio
source with flux 2.7~mJy at 1.4~Ghz by Miller \& Owen (2001; from work
by Sijbring 1993). 

The presence of UMBH in galaxies with stellar masses as low as that
of NGC\,1277 needs to be confirmed. This can be done by
spectroscopy of the stellar bulge at higher spatial resolution.
If the dust ring in NGC\,1277 is
accompanied by gas, as is likely, then the high expected circular
velocity ($\sim 480\kmps$) of that ring would provide convincing confirmation.
We shall proceed assuming that there {\em is} a UMBH in NGC\,1277.    

Van den Bosch et al (2012) list a further 5 galaxies which are
candidates for hosting ultramassive black holes, including 2 more in
the Perseus cluster (NGC\,1270 and UGC\,2689). NGC\,1270 is also in
the core of the cluster and lies in some of the deep Chandra
images. It is also characterised by a minicorona of temperature
1.07~keV plus a power-law component of 0.5--7\,keV luminosity $6.3\times
10^{40}\ergps$ (Sun et al 2007; Pearce et al 2012
submitted). NGC\,1270 is nearly 10 arcmin away from the Chandra
aimpoint so the PSF is too broad for an extent analysis. We note that
NGC\,4889, a BCG in the Coma cluster with a black hole of mass of
$2\times 10^{10}\Msun$ (McConnell et al 2011) also has a minicorona,
one of the first discovered (Vikhlinin et al 2001).

\section{DIscussion}
If we add the 6 candidate ultramassive black holes (UMBH) in the Table of van
den Bosch (2012) to the 2 found by McConnell et al (2012), then we
have 8 within about 105~Mpc. This represents a mass density of
$2\times 10^4 \Mpcc$ assuming a mean mass of $10^{10}\Msun$. If they
are all $2\times 10^{10}\Msun$ the mass density would be a factor
of two larger and if we
account for the fact that they are all Northern objects, which is the
minimum selection effect operating, then the mass density increases by
another factor of two. Their mass density 
could then easily be $10^{5}\Msun\Mpcc$ which is
a significant fraction of the mean mass density of black holes,
estimated by e.g.  Marconi et al (2004) at $4.6^{+1.9}_{-1.4}\times
10^5\Msun\Mpcc$ or Hopkins et al (2006) at $2.9^{+2.3}_{-1.2} \times
10^5\Msun\Mpcc$. The above 8 objects translate to a present UMBH  space
density  of at least $4\times 10^{-6}\Mpcc$. With three UMBH in the
Perseus cluster they appear to be strongly clustered.

The objects of McConnell et al (2012) are BCGs, 3 of the van den Bosch
(2012) sample are in the Perseus cluster and one other is in Abell
347. Hlavacek-Larrondo et al (2012) have indirect arguments for
UMBH in a distant population of BCGs.  There is
thus some evidence that the environment of a cluster, or whatever
turns into a cluster, is conducive to the growth of UMBH. The galaxies
that host them are of two types, the first are BCGs, with a high
stellar mass, the second are undistinguished early-type galaxies with
relatively low stellar mass, such as NGC\,1277.

\subsection{The growth of UMBH and their hosts}

If there is a UMBH in NGC\,1277 it is growing little at the present epoch, since
its current mass doubling time is many Hubble times. Its bolometric
Eddington limit is $3\times 10^{48}\ergps,$ which rivals the most
luminous objects known and requires an accretion rate of 100s
$\Msunpyr$. Even if it took billions of years to grow, we can expect
that it was a powerful quasar or blazar in the past. Hopkins et al (2006),
Ghisellini et al (2010) and Volonteri et al (2011) have all considered
the growth of black holes of mass exceeding $10^9\Msun$, based on
observations of quasars and blazars. The present number density they
estimate for black holes above that mass is less than we are inferring
for the UMBH, which are ten times more massive. This may indicate that
the UMBH, if real, are now too numerous to have been regular quasars
or blazars, but must have lost their accretion energy through
mechanical means, for example powerful winds, or been highly obscured.

A possible scenario then emerges where the enormous power of the
accreting UMBH stops any significant star formation in its
neighbourhood. Such extreme AGN feedback could have been too fierce
for new star formation to occur and only a compact stellar region or
bulge survives from the earliest stages of growth. Of course the black
hole has to be fuelled, which requires relatively low angular momentum
gas to be fed into the centre, more or less continuously. Dubois et al
(2012) show that this can occur at high redshifts. This would require
that feedback occurs in a bipolar fashion, as expected if jets or
winds from an accretion disc are involved, in order that the fuelling
along filaments can continue undisturbed. The large mass of the black
hole means that the orientation of the object will be relatively
fixed.

In this model, where the major UMBH growth phase is likely to happen
in deep potential wells before a redshift of 3, we are left with the
most massive black holes lying in compact bulges.  At redshift 3, most
UMBH would then resemble NGC\,1277.  (Van den Bosch et al 2012 have
already noted the resemblance between NGC\,1277 and typical red,
passive galaxies at earlier times.) What the host galaxy eventually
looks like then depends on how that bulge accretes more gas and stars,
and how further feedback shapes the gas and thus star formation. Such
a scenario is consistent with recent observational evidence that the
growth of massive galaxies from redshifts of 2.5 to 1 occurs in an
inside-out mode (van Dokkum et al 2010, Szomoru et al 2011 and
references therein).

If the UMBH host lies in a cluster then it only later accretes or
accumulates significant gas for star formation if it lies
at the centre of the potential well and so has a low velocity with
respect to its surroundings. Cannabalism, cooling flows and mergers
can supply that object with an extensive halo of stars and gas. The
object thus becomes a BCG.  If the UMBH bulge orbits in the core of a
cluster at $1000\kmps$, then it accumulates few stars and little gas
and could end up resembling NGC\,1277. Its outer dark matter halo will have
dissolved into the general cluster halo.

Note that there is a requirement in the model, which is not easy to
fulfill, that fuelling can proceed while powerful feedback is taking
place and we have appealed to a special geometry to allow this to
occur. This special requirement may be why few galaxies host a
UMBH. Only when the geometry of the fuelling filaments is appropriate
can a UMBH form, otherwise the central black hole is 10 to 100 times
less massive. 

The energy released in growing a black hole of mass $2\times
  10^{10}\Msun$ is $4\times 10^{63}\eta_{0.1}\erg,$ which will heat
  $10^{14}\Msun$ of gas, comparable to the gas in the present-day
  Perseus cluster, to $\sim 10\keV$. This could expel much of the gas
  in any early subcluster if the energy is widely distributed and
  little is radiated. Widespread heating to a few keV per particle by
  the growth of black holes has been invoked to explain the the X-ray
  properties of groups and clusters (e.g. the X-ray luminosity --
  temperature relation, Wu, Fabian \& Nulsen 2000), but 10\,keV is too
  much. Some inefficiency in the heating is required, such as could
  occur due to infrared radiation in an obscured scenario, or if the
  accretion process is intrinsically inefficient ($\eta<<0.1$).
  
\subsection{An alternative interpretation: a bottom-heavy IMF}

The data on NGC\,1277 presented by van den Bosch et al (2012) indicate
that it is an unusual galaxy with a high central mass-to-light
ratio. Future observations will confirm whether this is due to a
UMBH. An alternative possibility is that the Initial Mass Function
(IMF) of
the stars formed in the  core of the galaxy is very bottom heavy,
i.e. it has much of its mass in low mass stars which are not directly
detected. Recent work by van Dokkum \& Conroy (2010), Conroy \& van
Dokkum (2012) and Cappellari et al (2012) have
shown that the IMF in early type galaxies is increasingly skewed to
low mass stars as the velocity dispersion of the stars increases. This
could be due to the higher pressure of extended gaseous atmospheres in
deep potential wells (Krumholtz 2012). 

The high velocity dispersions found in the UMBH host galaxies already
imply a bottom-heavy IMF in the candidate UMBH hosts. It is possible
that the observations can be explained by a combination of low-mass
stars and a less massive black hole than a UMBH. van den Bosch et al
(2012) do however show that the mass-to-light ratio of the stellar
component needs to be exceptional (e.g. $M/L>10$) for the black hole
mass to be significantly reduced.

\section*{Acknowledgements}
We thank the referee for helpful comments.
We thank the Chandra X-ray Observatory team for taking the observations.
The image of NGC\,1277 used in Fig.~4 is from the Mikulski Archive for
Space Telescopes (MAST). ACF thanks the Royal Society for support.


\end{document}